%% file: main.tex
\documentclass[sigplan,nonacm]{acmart}
\usepackage{booktabs}
\usepackage{subcaption}
\usepackage{xspace}
\usepackage{listings}
\usepackage{xcolor}
\usepackage{tikz}
\usetikzlibrary{positioning,arrows.meta,fit,backgrounds,patterns,calc,shapes.geometric}

\usepackage[most]{tcolorbox}
\usepackage{float}       
\usepackage{ragged2e}    

\definecolor{pastelblue}{RGB}{173,216,230}
\definecolor{pastelyellow}{RGB}{255, 255, 179}

\tcbset{textmarker/.style={%
    enhanced,
    parbox=false,boxrule=0mm,boxsep=0mm,arc=0mm,
    outer arc=0mm,left=4mm,right=3mm,top=7pt,bottom=7pt,
    toptitle=1mm,bottomtitle=1mm,oversize}}

\newtcolorbox{sidebar_box}[2][]{textmarker,
    borderline west={6pt}{0pt}{pastelblue},
    colback=pastelblue!10!white,
    halign=justify,
    title=\textcolor{black}{\textbf{#2}},
    title code={
      \path[fill=pastelblue!10!white] (title.south west) rectangle (title.north east);
      \path[draw=pastelblue,solid,line width=0.75mm]
      ([xshift=0mm]title.south west) -- ([xshift=0mm]title.south east);
    },
    #1
}

\usepackage{pgfplots}
\pgfplotsset{compat=1.18}
\usepackage{algorithm}
\usepackage{algorithmic}
\usepackage{multirow}
\usepackage{pifont}

\newcommand{\sysname}{\textsc{Fleet}\xspace}
\newcommand{\xcdtask}{Chiplet-task\xspace}

\begin{document}

\setlength{\headheight}{13pt}
\fancypagestyle{standardpagestyle}{%
  \fancyhf{}%
  \fancyfoot[C]{\thepage}%
  \renewcommand{\headrulewidth}{0pt}%
}
\pagestyle{standardpagestyle}

\title{\sysname: Hierarchical Task-based Abstraction for Megakernels on Multi-Die GPUs}

\renewcommand{\shortauthors}{Chowdhary et al.}

\author{Sangeeta Chowdhary}
\affiliation{\institution{AMD Research}\country{}}

\author{Ryan Swann}
\affiliation{\institution{AMD Research}\country{}}

\author{Sean Siddens}
\affiliation{\institution{AMD Research}\country{}}

\author{Muhammad Osama}
\affiliation{\institution{AMD Research}\country{}}

\author{Stephen Neuendorffer}
\affiliation{\institution{AMD Research}\country{}}

\author{Alexandru Dutu}
\affiliation{\institution{AMD Research}\country{}}

\author{Karthik Sangaiah}
\affiliation{\institution{AMD Research}\country{}}

\author{Sandeepa Bhuyan}
\affiliation{\institution{AMD Research}\country{}}

\author{Samuel Bayliss}
\affiliation{\institution{AMD Research}\country{}}

\author{Ganesh Dasika}
\affiliation{\institution{AMD Research}\country{}}

\thanks{Correspondence to: Sangeeta Chowdhary \texttt{<sangeeta.chowdhary@amd.com>}}

\begin{abstract}
Modern GPUs adopt chiplet-based designs with multiple private cache hierarchies, but current programming models (CUDA/HIP) expose a flat execution hierarchy that cannot express chiplet-level locality or synchronization. This mismatch leads to redundant memory traffic and poor cache utilization in memory-bound workloads such as LLM inference.

We present \sysname, a multi-level task model that maps computation to memory scopes. \sysname introduces Chiplet-tasks, a new abstraction that binds work and data to a chiplet and enables coordination through its shared L2 cache. Wavefront-, CU-, and device-level tasks align with existing abstractions, while Chiplet-tasks expose a previously unaddressed level of the hierarchy. \sysname is implemented as a persistent kernel runtime with per-chiplet scheduling, allowing workers within a chiplet to cooperatively execute tasks with coordinated cache reuse.
On AMD Instinct MI350 with Qwen3-8B, \sysname achieves 1.3--1.5$\times$ lower decode latency than vLLM at batch sizes 1--8 through persistent kernel execution and per-chiplet scheduling. At larger batch sizes, cooperative weight tiling increases L2 hit rate (from 12\% to 54\% at batch size 32 and from 39\% to 61\% at batch size 64), reducing HBM traffic by up to 37\% and delivering 1.27--1.30$\times$ speedup over a chiplet-unaware megakernel baseline.
\end{abstract}

\maketitle

\section{Introduction}
\label{sec:intro}

Modern GPU architectures have moved toward chiplet-based, multi-die designs as a way of continuing to improve compute density.
For instance, AMD's Instinct MI300X~\cite{AMD:2024:AIM} and MI350/MI355 implement eight XCDs (Accelerator Complex Dies) with private 4\,MB L2 caches while NVIDIA's Blackwell integrates two dies on a single package~\cite{NVIDIA:2024:NBA}.
This multi-die approach is critical for scaling AI compute, enabling more densely connected accelerators without scale-out networking.
Yet the CUDA/HIP execution model (\emph{thread $\rightarrow$ wavefront $\rightarrow$ workgroup $\rightarrow$ grid}) has not adapted to this architectural shift.

In particular, there is no direct way to express data affinity between
groups of workgroups, or to scope work to a specific chiplet's memory hierarchy. Both AMD and NVIDIA face this challenge, yet their approaches differ substantially (e.g., chiplet swizzling~\cite{Hu:2025:HFA}), making platform-specific workarounds non-portable. The resulting gap is especially acute for LLM inference, where interactive chatbots, coding assistants, and voice agents demand sub-10\,ms per-token decode~\cite{Dao:2024:FFA3}, a regime dominated by memory bandwidth.
  
One response is ``persistent kernel'' programming~\cite{Spector:2025:LMN, Nrusimha:2025:FWM}, where a kernel dispatches a grid of workgroups that matches the hardware dimensions and persists for the lifetime of the computation, rather than launching separate kernels per operator.

\input{figures/mem_hierarchy}

This paper presents \sysname, a multi-level task model that maps operators to memory scopes matching their working sets.
In \sysname, an application workload is partitioned into \emph{tasks} with explicit dependencies between them.
Tasks are scoped to different levels of the memory hierarchy:
\emph{wavefront-tasks} operate within a single wavefront using registers and LDS;
\emph{CU-tasks} occupy a single compute unit, communicating through the L1/LDS hierarchy;
\emph{Chiplet-tasks} span all workers on a chiplet, coordinating data access through the XCD-local L2 cache;
and a \emph{device-task}, comprised of eight Chiplet-tasks, collectively executes one operator across the entire device.
When compared with CUDA/HIP, wavefront-tasks, CU-tasks, and device-tasks roughly correspond to familiar concepts of wavefronts, workgroups, and grids by design. The key new abstraction is the Chiplet-task, which enables the programmer to express behavior that is usefully scoped to chiplet boundaries and the corresponding L2 memory hierarchy. In this work, we use CU-tasks and Chiplet-tasks for compute operations.

\sysname is implemented as a persistent kernel device-side runtime with per-chiplet scheduling.
One workgroup per chiplet is designated as a \emph{scheduler}; the rest are \emph{workers}.
Each scheduler retrieves task descriptors from a data structure in device memory and distributes these tasks to its local workers on the same chiplet.
For Chiplet-tasks, the scheduler broadcasts the same task to all workers in a single chiplet, which can then cooperatively execute the workload with coordinated L2 reuse.
For CU-tasks and wavefront-tasks, each scheduler dispatches individual tasks round-robin among the CUs in the same chiplet.
This hierarchical dispatch leverages chiplet-local L2 coherence for low-overhead communication within a Chiplet-task, reducing the need for more expensive cross-chiplet synchronization.


\section{Background and Motivation}
\label{sec:background}
The CUDA/HIP programming model was designed for monolithic GPUs, in which all CUs share a single unified L2 cache.
Each level of the hardware memory hierarchy is mapped to a programming abstraction with a well-defined scope, as shown in Table~\ref{tab:memory-scope}.
On a monolithic GPU, any CU could access any L2 line with the same latency and bandwidth, so the programmer never needed to reason about \emph{where} data was cached.              
\begin{table}[t]
\centering
\caption{Memory hierarchy to programming model mapping on monolithic GPUs. On chiplet GPUs, the L2 scope breaks: it is physically partitioned across dies, but the programming model still treats it as device-scoped.}
\label{tab:memory-scope}
\small
\begin{tabular}{@{}llll@{}}
\toprule
\textbf{Compute}    & \textbf{Memory}    & \textbf{CUDA/HIP} & \textbf{\sysname} \\
\textbf{Hierarchy} & \textbf{Hierarchy} & \textbf{Abstraction} & \textbf{Abstraction} \\
\midrule
SIMD & Register file      & Wavefront             & Wavefront-task  \\
CU/SM & LDS  & Workgroup    & CU-task               \\
Chiplet & L2 cache           & ---             & Chiplet-task                \\
Device & HBM                & Device         & Device-task        \\
\bottomrule
\end{tabular}
\end{table}

Widespread demand for larger models and higher throughput has driven GPU vendors to scale chip size beyond what monolithic designs can deliver. NVIDIA's Blackwell architecture partitions a device into 2 chiplets each with a local L2 cache partition~\cite{Jarmusch:2025:MNB}.  A globally coherent L2 cache is implemented across the L2 cache partitions.   On the other hand, AMD's MI300X and MI350 are implemented using eight chiplets (XCDs), each with a private, non-coherent 4\,MB L2 cache as shown in Figure~\ref{fig:mi300x-arch}.  In either case, memory latency to data is non-uniform and will be lower if data is resident in the same chiplet, but there is not a direct way to view the device as a single CUDA/HIP device while controlling which workgroups are localized to which XCDs.

\emph{This architectural shift introduces NUMA-like programming effects within a single device which were previously only present between devices.}  As a result, although memory bandwidth is abundant in these GPUs (5+\,TB/s), L2 cache locality becomes a more important optimization goal.  This is particularly acute in the AMD architecture, in which workgroups scheduled to different XCDs cannot share L2 data, resulting in increased L2 cache contention within an XCD and wasted bandwidth.

%

We first detail AMD's chiplet architecture and cache coherence model (\S\ref{sec:chiplet-arch}), then characterize why LLM decode is memory-bound (\S\ref{sec:llm-decode}), empirically demonstrate the performance cost of chiplet-unaware scheduling, and finally discuss limitations of prior graph-based approaches (\S\ref{sec:graphs}).

\subsection{Chiplet GPU Architecture}
\label{sec:chiplet-arch}

In this paper, we focus and evaluate our results on AMD's Instinct MI350X architecture\footnote{We did not have access to a NVIDIA Blackwell GPU.}. AMD's MI300 series architectures (MI300X, MI350X, MI355X) integrate eight XCDs on a single package via advanced packaging.
MI350 delivers 256 compute units (CUs) across 8 XCDs (32 CUs per XCD) with HBM3 bandwidth exceeding 5\,TB/s.
The devices also have 8 stacks of HBM (totaling 288 GB in MI350), which is visible in a single global address space, allowing workgroups running on any chiplet to access data in any HBM stack.
Data is cached on-chip with both per-CU L1 cache and a per-XCD L2 cache.  However unlike typical monolithic GPUs where all CUs share a unified L2 cache, the L2 cache in these GPUs is \emph{partitioned}: each XCD has a private 4\,MB L2 cache, for a total of 32\,MB across the package.
In addition to the L2 cache, both MI300X and MI350 include a 256\,MB last-level cache (LLC), also called the Infinity Cache, which is shared across all eight XCDs.
The LLC sits between L2 and HBM, serving as a victim cache for L2 evictions.  

When a HIP kernel executes, the workgroups in a grid are typically distributed across different XCDs by the hardware scheduler, allowing the grid to leverage all 32 MB of L2 cache.
This workload distribution is effective when workgroups access independent data, but can be less efficient when workgroups access the same data from different chiplets.  

\paragraph{L2 cache coherence and scope control.}
When a CU on XCD~$i$ issues a global memory load, the request first checks XCD~$i$'s local L2 cache.
On a miss, the request traverses the Infinity Fabric to the IO die, where the request may be routed to HBM or to a peer XCD holding a dirty copy.
The specific behavior on CDNA3/CDNA4 is governed by two scope bits (\texttt{SC1}, \texttt{SC0}) and a non-temporal bit (\texttt{NT}) on each memory instruction~\cite{AMD:2025:CDNA4ISA}.
\texttt{SC1} and \texttt{SC0} encode the coherence scope: \texttt{0\_0}{=}wave, \texttt{0\_1}{=}group, \texttt{1\_0}{=}device, \texttt{1\_1}{=}system.
Default loads and stores use \texttt{SC1{=}0, SC0{=}0, NT{=}0} (wave scope), which caches data in both L1 (TCP, read-only) and L2 with an LRU policy.
With wave scope, the L2 treats data as local to the XCD: no cross-XCD coherence probes are issued, and stale copies on peer XCDs are not invalidated.
Making data produced by XCD~$i$ visible to XCD~$j$ requires explicit software action: the producer must issue \texttt{buffer\_wbl2} to write back dirty L2 lines, and the consumer must invalidate stale L2 entries.
Operations at device or system scope (\texttt{SC0{=}1} or \texttt{SC1{=}1}) engage the coherence infrastructure: L2 uses a \emph{Coherent\_Cache\_Bypass} policy that writes back modified lines before reissuing reads, ensuring cross-XCD visibility at the cost of higher latency.


\begin{sidebar_box}{Implication for LLM Decode}
The practical consequence for LLM decode is that weight data, loaded via default instructions, is cached independently in each XCD's L2 with no sharing. If two CUs on different XCDs read the same weight row, the data is fetched from HBM (or LLC) twice and cached in two separate L2 partitions. This leads to redundant memory traffic and reduced effective bandwidth utilization.
\end{sidebar_box}

\begin{table}[t]
\centering
\caption{Characterization of chiplet-unaware LLM decode on MI350. Linear operations dominate compute time and have poor L2 utilization.}
\label{tab:motivation}
\footnotesize
\begin{tabular}{|l|r|r|}
\hline
\textbf{Metric} & \textbf{Linear} & \textbf{Attention} \\
\hline
\% of decode time & 95\% & 5\% \\
\hline
Weight working set / layer & 368\,MB & --- \\
\hline
Weight per XCD (uniform) & 46\,MB & --- \\
\hline
XCD L2 cache capacity & 4\,MB & 4\,MB \\
\hline
L2 hit rate (bs=1, no coop.) & 16.4\% & --- \\
\hline
Cycles per task & $\sim$104K & $\sim$3.8K \\
\hline
\end{tabular}
\end{table}

\subsection{LLM Decode is Memory and Kernel Launch Latency Bound}
\label{sec:llm-decode}

LLM inference consists of two phases: \emph{prefill} (processing the input prompt, compute-bound) and \emph{decode} (generating tokens autoregressively, memory-bound)~\cite{Dao:2022:FFA, Pope:2023:EST}.
Decode dominates serving latency \cite{Patel:2024:SEG, Jiang:2026:PAD} and is the focus of this work.

Each decode step for a transformer layer performs:
\begin{enumerate}
\item RMSNorm~\cite{Zhang:2019:RMS} (element-wise, negligible cost)
\item QKV projection: $\mathbf{x} \cdot \mathbf{W}_{qkv}$, where $\mathbf{W}_{qkv} \in \mathbb{R}^{d \times 3d'}$
\item Multi-head attention with KV cache
\item Output projection: $\mathbf{x} \cdot \mathbf{W}_o$
\item RMSNorm + gate-up projection: $\mathbf{x} \cdot \mathbf{W}_{gu}$, where $\mathbf{W}_{gu} \in \mathbb{R}^{d \times 2d_{ff}}$
\item SiLU activation $\times$ element-wise multiply
\item Down projection: $\mathbf{x} \cdot \mathbf{W}_{down}$ + residual add
\end{enumerate}

For Qwen3-8B~\cite{Qwen:2025:QTR} ($d{=}4096$, $d_{ff}{=}12288$), the weight matrices total 368\,MB per layer in bf16.
At a batch size of 128, the activation matrices are $128 \times 4096 \times 2 = 1$\, MB, which is three orders of magnitude smaller than the weights. 
Hence, we focus on optimizations that reduce the external memory bandwidth used to load weights.
The L2 cache could serve weight data at a much higher bandwidth than HBM (${\sim}$100\, TB/s aggregate across 8 XCDs vs.\ 5.3\, TB/s for HBM). Still, a standard serving system executes each operation as a separate GPU kernel, resulting in almost 250 launches per decode token across 36 layers.
Between launches, the L2 cache contents from one operator are not preserved for the next to maintain coherence at device synchronization barriers, forcing each kernel to reload its weights and activations from HBM independently.
This motivates fusing all operators into a single persistent kernel that retains L2 state across operators.

\subsection{Limitations of CUDA/HIP Graph Capture}
\label{sec:graphs}
CUDA and HIP graph capture~\cite{NVIDIA:2024:CCT} reduces kernel launch overhead by recording and replaying kernel sequences. In practice, serving frameworks such as vLLM and SGLang capture a separate graph for each batch size, because a captured graph bakes in kernel arguments, grid dimensions, and memory pointers; mismatched requests fall back to eager execution, causing multi-$\times$ latency spikes.

Beyond batch-size specialization, graphs inherit structural limitations that persistent megakernels do not~\cite{Spector:2025:LMN, Cheng:2025:MPK, Nrusimha:2025:FWM}: dependent nodes still execute sequentially at kernel-scope with inter-kernel gaps, residual launch overhead (reduced but not eliminated), and no cross-kernel data reuse (kernel boundaries flush L2, forcing intermediates through HBM).

\sysname avoids these trade-offs: as a persistent kernel, it pays the launch cost exactly once and delivers deterministic latency across all batch sizes.

\begin{figure*}[t]
\centering
\resizebox{\textwidth}{!}{%
\begin{tikzpicture}[
    xcdlbl/.style={font=\normalsize\bfseries, anchor=south},
    wkr/.style={draw, minimum width=0.5cm, minimum height=0.5cm, inner sep=0pt, font=\small\bfseries},
    tile/.style={draw, minimum width=0.7cm, minimum height=0.55cm, inner sep=0pt, font=\small},
    annot/.style={font=\small, text=black!70},
    >=Stealth
]

\node[font=\large\bfseries, anchor=south] at (2.4, 5.8) {(a) Standard Scheduling};

\foreach \x/\lbl in {0/0, 2.4/1, 4.8/2} {
    \node[draw, rounded corners=3pt, minimum width=1.8cm, minimum height=1.2cm, fill=blue!5, thick] (sxcd\lbl) at (\x, 4.4) {};
}
\node[xcdlbl] at (0, 5.1) {XCD 0};
\node[xcdlbl] at (2.4, 5.1) {XCD 1};
\node[xcdlbl] at (4.8, 5.1) {XCD 2};

\node[wkr, fill=red!30] at (-0.35, 4.6) {};
\node[wkr, fill=green!30] at (0, 4.2) {};
\node[wkr, fill=blue!30] at (0.35, 4.6) {};
\node[wkr, fill=blue!30] at (2.05, 4.6) {};
\node[wkr, fill=red!30] at (2.4, 4.2) {};
\node[wkr, fill=green!30] at (2.75, 4.6) {};
\node[wkr, fill=green!30] at (4.45, 4.6) {};
\node[wkr, fill=blue!30] at (4.8, 4.2) {};
\node[wkr, fill=red!30] at (5.15, 4.6) {};

\node[annot, anchor=east, align=right] at (-1.0, 2.0) {Weight\\[-2pt]matrix};
\node[tile, fill=red!20] at (-0.4, 2.3) {};
\node[tile, fill=blue!20] at (0.4, 2.3) {};
\node[tile, fill=green!20] at (-0.4, 1.7) {};
\node[tile, fill=red!20] at (0.4, 1.7) {};

\node[tile, fill=green!20] at (2.0, 2.3) {};
\node[tile, fill=red!20] at (2.8, 2.3) {};
\node[tile, fill=blue!20] at (2.0, 1.7) {};
\node[tile, fill=green!20] at (2.8, 1.7) {};

\node[tile, fill=blue!20] at (4.4, 2.3) {};
\node[tile, fill=green!20] at (5.2, 2.3) {};
\node[tile, fill=red!20] at (4.4, 1.7) {};
\node[tile, fill=blue!20] at (5.2, 1.7) {};

\draw[->, thick, gray!60] (0, 3.7) -- (0, 2.65);
\draw[->, thick, gray!60] (2.4, 3.7) -- (2.4, 2.65);
\draw[->, thick, gray!60] (4.8, 3.7) -- (4.8, 2.65);

\def\rx{9.0}
\node[font=\large\bfseries, anchor=south] at (\rx+2.4, 5.8) {(b) \sysname};

\node[draw, rounded corners=3pt, minimum width=1.8cm, minimum height=1.2cm, fill=red!15, thick] at (\rx+0, 4.4) {};
\node[draw, rounded corners=3pt, minimum width=1.8cm, minimum height=1.2cm, fill=blue!15, thick] at (\rx+2.4, 4.4) {};
\node[draw, rounded corners=3pt, minimum width=1.8cm, minimum height=1.2cm, fill=green!15, thick] at (\rx+4.8, 4.4) {};
\node[xcdlbl] at (\rx+0, 5.1) {XCD 0};
\node[xcdlbl] at (\rx+2.4, 5.1) {XCD 1};
\node[xcdlbl] at (\rx+4.8, 5.1) {XCD 2};

\node[wkr, fill=red!30] at (\rx-0.35, 4.6) {};
\node[wkr, fill=red!30] at (\rx+0, 4.2) {};
\node[wkr, fill=red!30] at (\rx+0.35, 4.6) {};
\node[wkr, fill=blue!30] at (\rx+2.05, 4.6) {};
\node[wkr, fill=blue!30] at (\rx+2.4, 4.2) {};
\node[wkr, fill=blue!30] at (\rx+2.75, 4.6) {};
\node[wkr, fill=green!30] at (\rx+4.45, 4.6) {};
\node[wkr, fill=green!30] at (\rx+4.8, 4.2) {};
\node[wkr, fill=green!30] at (\rx+5.15, 4.6) {};


\node[tile, fill=red!25] at (\rx-0.4, 2.3) {};
\node[tile, fill=red!25] at (\rx+0.4, 2.3) {};
\node[tile, fill=red!25] at (\rx-0.4, 1.7) {};
\node[tile, fill=red!25] at (\rx+0.4, 1.7) {};

\node[tile, fill=blue!25] at (\rx+2.0, 2.3) {};
\node[tile, fill=blue!25] at (\rx+2.8, 2.3) {};
\node[tile, fill=blue!25] at (\rx+2.0, 1.7) {};
\node[tile, fill=blue!25] at (\rx+2.8, 1.7) {};

\node[tile, fill=green!25] at (\rx+4.4, 2.3) {};
\node[tile, fill=green!25] at (\rx+5.2, 2.3) {};
\node[tile, fill=green!25] at (\rx+4.4, 1.7) {};
\node[tile, fill=green!25] at (\rx+5.2, 1.7) {};


\draw[->, thick, red!60] (\rx+0, 3.7) -- (\rx+0, 2.65);
\draw[->, thick, blue!60] (\rx+2.4, 3.7) -- (\rx+2.4, 2.65);
\draw[->, thick, green!50!black] (\rx+4.8, 3.7) -- (\rx+4.8, 2.65);

\draw[thick, dashed, gray!40] (7.0, 0.8) -- (7.0, 6.2);

\end{tikzpicture}%
}
\caption{Standard block scheduling (a) assigns workers on the same XCD to different GEMMs, thrashing L2. \sysname's scheduling (b) partitions weight columns across XCDs and coordinates all workers on each XCD to read the same partition, converting L2 misses into hits.}
\label{fig:hero}
\end{figure*}

\section{The \sysname Task Model}
\label{sec:model}

This section presents the \sysname task model from abstractions to mechanisms.
We first define the four-level task hierarchy that maps operators to chiplet resource boundaries (\S\ref{sec:hierarchy}).
With the hierarchy defined, we describe cooperative tiling, the traversal order that converts L2 misses into hits (\S\ref{sec:coop-tiling}).
Finally, we present the hierarchical synchronization protocol, which makes cross-task dependencies efficient on partitioned L2 caches (\S\ref{sec:two-level-events}).

\subsection{Multi-Level Task Hierarchy}
\label{sec:hierarchy}

The fundamental design principle of \sysname is that tasks should be written at the \emph{hardware granularity which matches the operator's resource needs}.
Modern chiplet GPUs have a deep resource hierarchy (registers, shared memory, L2 cache, HBM), and each level has different capacity, bandwidth, and sharing characteristics.
\sysname exposes this hierarchy through four task levels (Table~\ref{tab:hierarchy}):

\begin{table}[t]
\centering
\caption{Multi-level task hierarchy in \sysname. Each level maps to a hardware resource boundary with specific memory scope and typical operators.}
\label{tab:hierarchy}
\small
\resizebox{\columnwidth}{!}{%
\begin{tabular}{llllr}
\toprule
\textbf{Level} & \textbf{HW Scope} & \textbf{Memory} & \textbf{Typical Op} & \textbf{Workers} \\
\midrule
Wavefront-task  & 1 wavefront   & Regs, LDS        & SiLU, res.\ add   & 1 \\
CU-task    & 1 block  & LDS, L2          & Attn, RMSNorm     & 1 \\
Chiplet-task   & 1 XCD    & L2, HBM          & GEMM partition    & 31 \\
Device-task      & 8 XCDs   & Global HBM       & Full GEMM/attn    & 248 \\
\bottomrule
\end{tabular}%
}
\label{fig:task-hierarchy}
\end{table}

\paragraph{wavefront-tasks} are the lightest units of work, executing within a single wavefront (64 threads on AMD).
Element-wise operators naturally map to wavefront-tasks: SiLU activation, element-wise multiply, residual addition, and RoPE.

\paragraph{CU-tasks} occupy a single compute unit (workgroup), using LDS for intra-block communication and L2 for data access.
Operators with moderate data requirements map to CU-tasks: individual attention heads, RMSNorm, and argmax/sampling.

\paragraph{Chiplet-tasks} span all workers on a single XCD (31 CUs out of 32 on MI350), with an explicit \emph{L2 cache budget}. 
This is the key new abstraction: the programmer specifies the data partition, tiling, and L2 working set for each chiplet.
GEMM operations typically map to Chiplet-tasks, with weight matrices partitioned so that each XCD processes a cache-friendly slice.

\paragraph{Device-tasks} coordinate 8 Chiplet-tasks (one per chiplet) to collectively execute one operator.
Device-tasks have barrier semantics: a device-task completes only when all 8 Chiplet-tasks finish.
Each XCD writes its columns at a strided offset within a shared output buffer, assembling the result in place without reduction. For operators that partition along the reduction dimension (e.g., split-K GEMM), the device-task can include a reduction phase after the chiplet-tasks complete.

\paragraph{Task Dependence}
Inter-task dependencies are expressed as \emph{events}: each task declares which event it signals on completion and which events it waits for.
Because Chiplet-tasks group all workers on a chiplet into a single unit, only one event is needed per chiplet per dependency edge, rather than one per worker. This hierarchical representation reduces the number of cross-chiplet synchronization events by $W\times$ (where $W$ is workers per chiplet), a benefit we quantify in \S\ref{sec:two-level-events}.

\section{Chiplet-aware GPU Optimization} 
\label{sec:optimization}

Chiplet-tasks give control over what data each chiplet accesses.
This section demonstrates how to exploit that control to increase L2 cache utilization.
We focus on a scenario central to LLM inference: linear layer GEMMs in dense models~(\S\ref{sec:coop-tiling}).

\subsection{Optimizing Batched GEMV}
\label{sec:coop-tiling}

Batched matrix operations can provide copious amounts of data reuse since each weight is reused among multiple elements of the batch.  
Using \sysname, we can express this locality by partitioning the execution of a single linear layer into multiple Chiplet-tasks.
For an [M,K] * [K,N] GEMM, we partition the output matrix column-wise so that each Chiplet-task computes an independent [M,K] * [K,N/8] = [M,N/8] sub-matrix. 

For typical operation sizes (such as the gate-up projection in the Qwen3-8B FFN, where each XCD's partition is $[4096, 3072]$ = 24\,MB in bf16), the weight matrices don't completely fit in the 4\,MB L2 cache.
In order to ensure effective temporal reuse, we apply \emph{windowed M-major traversal} (inspired by HipKittens Algorithm~1~\cite{Hu:2025:HFA}) as shown in Figure~\ref{fig:tiling}. This approach attempts to ensure that different CUs accessing the same weights do so in the same order, helping the L2 cache effectively reduce load bandwidth to external memory.

In this scheme, within each \xcdtask, each worker will compute a tile [m,n] of the output completely using [m, 0:K] tiles of activations and [0:K, n] tiles of weights before moving on to the next tile.
Conceptually, we attempt to use all of the [0:K, 0] weights first, then process all for the [0:K, 1] weights, etc.

We evaluate two strategies for distributing M-tiles (batch partitions) across XCDs:

\paragraph{M-tile traversal.}
All 8 XCDs receive the same set of $m\_tiles = \lceil B/T_M \rceil$ M-tiles, where $B$ is the batch size and $T_M$ is the GEMM tile height along the batch dimension.
Within each XCD, workers iterate over M-tiles before advancing to the next weight column (N-tile).
This means multiple workers on the same XCD process the same weight column at roughly the same time: the first worker loads the weight tile from HBM into L2, and subsequent workers hit L2.
With $R = \min(W, m\_tiles)$ workers sharing each weight tile, where $W$ is the number of workers per XCD, the expected L2 hit rate for weights is $(R-1)/R$.
If N-major ordering were used instead, consecutive workers would each read a different weight tile, preventing L2 reuse.

\paragraph{M-split traversal.}
Each XCD is assigned a disjoint M-tile: XCD~$k$ processes M-tile $k \bmod m\_tiles$.
When $m\_tiles < 8$, multiple XCDs share the same M-tile, but each XCD's workers process disjoint weight columns with no cooperative weight sharing.
This isolates the benefit of Chiplet-task scheduling from L2 cooperative tiling.

Note that if N-major ordering were used instead, consecutive workers would each read a \emph{different} tile of weights.  The resulting weights would take longer to load from external memory, and then would likely be evicted from the L2 cache as other rows of weights were required.

\paragraph{Cache modifier policy.}
On CDNA3/CDNA4 AMD GPUs, each memory instruction carries scope and non-temporal (NT) bits that control L2 allocation policy.
Specifically, the streaming behavior (using cache modifier bits \texttt{sc1=1, nt=1}) instructs the L2 to temporarily allocate data, enabling short-lived reuse, but to mark the line for immediate eviction when the cache needs space.
This contrasts with the default cache all behavior, which uses standard LRU eviction and allows transient data to displace more valuable resident data. We exploit these modifiers with a three-tier cache policy:
\begin{enumerate}

\item \textbf{Weight loads}: cache-streaming (\texttt{sc1=1, nt=1}).
Weight tiles are read once per GEMM and do not need to persist across layers, making them a natural fit for the streaming cache policy.
Cooperative M-tile sharing still generates L2 hits because consecutive workers access the same weight column within a short temporal window, before the streaming policy evicts the line.

\item \textbf{Activation stores}: non-temporal (\texttt{NT=1}), bypassing L2 so that transient GEMM outputs, RMSNorm results, and SiLU activations do not evict weight tiles during cooperative execution.
\item \textbf{Scheduler communication}: non-temporal loads for cross-XCD event polling, bypassing stale L2 copies to read fresh values from HBM; intra-XCD communication uses volatile loads through the shared L2.
\end{enumerate}

\input{figures/xcd_tiling}

\paragraph{Operator fusion and L2 residency.}
Fusing operators within a single Chiplet-task eliminates intermediate buffer writes that would otherwise flush the dirty writes from the L2 cache. We evaluated fusing the SiLU activation into the gate+up GEMM Chiplet-task, eliminating the separate SiLU task and its output buffer. This improves L2 hit rate from 9.4\% to 17.4\% at bs=1 for \sysname. Notably, fusing SiLU into a CU-task yields a comparable improvement (8.1\% to 16.5\%), indicating that the L2 benefit comes from eliminating intermediate buffer traffic rather than from XCD-scoped coordination. The fusion reduces one buffer write and one buffer read per layer, keeping activation data in registers/LDS rather than cycling through L2.

\paragraph{N-split vs.\ K-split across XCDs.}
\label{sec:nsplit-vs-ksplit}
\sysname partitions GEMM output columns across XCDs (\emph{N-split}), where each XCD computes an independent $[M, N/8]$ slice with full K reduction and no cross-XCD synchronization.
An alternative strategy partitions the reduction dimension (\emph{K-split})~\cite{Osama:2023:SWP}: within a single XCD, multiple waves in a workgroup each handle $K/k$ of the reduction, then merge partial results via intra-wave register shuffles and cross-wave LDS reduction, entirely on-chip with no atomics or global memory traffic.
The two strategies are complementary: N-split with cooperative L2 management dominates at small batch sizes (bs=1--16) where persistent kernel scheduling overhead is the bottleneck, while K-split with wave-level reduction excels at bs$\geq$32 where full CU utilization and auto-tuned GEMM tiles are critical.

\section{System Design}
\label{sec:system}

The previous section defined the task model, what tasks are, how they are dispatched, and how they synchronize.
This section describes the system that executes this model: the compiler that generates task descriptors, the persistent kernel runtime, and the code generation pipeline.

\begin{figure*}[t]
\centering
\resizebox{\textwidth}{!}{%
\begin{tikzpicture}[
    sched/.style={draw, rounded corners=2pt, minimum width=0.7cm, minimum height=0.5cm, fill=purple!25, font=\scriptsize\bfseries, inner sep=1pt},
    wkr/.style={draw, rounded corners=2pt, minimum width=0.5cm, minimum height=0.5cm, fill=blue!15, font=\scriptsize\bfseries, inner sep=1pt},
    xcdbox/.style={draw, dashed, rounded corners=5pt, inner sep=6pt, thick, blue!30},
    tasknode/.style={draw, rounded corners=2pt, minimum width=1.6cm, minimum height=0.45cm, font=\footnotesize\bfseries, inner sep=2pt},
    cu/.style={tasknode, fill=green!20},
    xcd/.style={tasknode, fill=blue!15, thick},
    wavefront/.style={tasknode, fill=yellow!25},
    arr/.style={->, >=Stealth, thick, black!50},
    annot/.style={font=\footnotesize, text=black!70},
    >=Stealth
]

\node[font=\normalsize\bfseries, anchor=south] at (3.0, 10.8) {(a) Task Graph: 1 Transformer Layer (bs=1)};

\def\lx{0}
\node[font=\small\bfseries, text=black!60] at (\lx+1.0, 10.3) {Standard (1,407 tasks)};

\node[cu] (s-rms1) at (\lx+1.0, 9.6) {RMSNorm};
\node[annot, anchor=west] at (\lx+2.0, 9.6) {\scriptsize 1 task};

\node[cu] (s-qkv) at (\lx+1.0, 8.7) {QKV Linear};
\node[annot, anchor=west] at (\lx+2.0, 8.7) {\scriptsize 96 tasks};

\node[cu] (s-attn) at (\lx+1.0, 7.8) {Attention};
\node[annot, anchor=west] at (\lx+2.0, 7.8) {\scriptsize 8+8 tasks};

\node[cu] (s-oproj) at (\lx+1.0, 6.9) {O-Proj+Res};
\node[annot, anchor=west] at (\lx+2.0, 6.9) {\scriptsize 256 tasks};

\node[cu] (s-rms2) at (\lx+1.0, 6.0) {RMSNorm};
\node[annot, anchor=west] at (\lx+2.0, 6.0) {\scriptsize 1 task};

\node[cu] (s-gate) at (\lx+1.0, 5.1) {Gate+Up};
\node[annot, anchor=west] at (\lx+2.0, 5.1) {\scriptsize 192 tasks};

\node[wavefront] (s-silu) at (\lx+1.0, 4.2) {SiLU $\times$ mul};
\node[annot, anchor=west] at (\lx+2.0, 4.2) {\scriptsize 96 tasks};

\node[cu] (s-down) at (\lx+1.0, 3.3) {Down+Res};
\node[annot, anchor=west] at (\lx+2.0, 3.3) {\scriptsize 256 tasks};

\draw[arr] (s-rms1) -- (s-qkv);
\draw[arr] (s-qkv) -- (s-attn);
\draw[arr] (s-attn) -- (s-oproj);
\draw[arr] (s-oproj) -- (s-rms2);
\draw[arr] (s-rms2) -- (s-gate);
\draw[arr] (s-gate) -- (s-silu);
\draw[arr] (s-silu) -- (s-down);

\def\rx{5.5}
\node[font=\small\bfseries, text=black!60] at (\rx+1.0, 10.3) {\sysname (543 tasks)};

\node[cu] (f-rms1) at (\rx+1.0, 9.6) {RMSNorm};
\node[annot, anchor=west] at (\rx+2.0, 9.6) {\scriptsize 1 CU-task};

\node[xcd] (f-qkv) at (\rx+1.0, 8.7) {QKV Proj};
\node[annot, anchor=west] at (\rx+2.0, 8.7) {\scriptsize 8 Chiplet-tasks};

\node[cu] (f-attn) at (\rx+1.0, 7.8) {Attention};
\node[annot, anchor=west] at (\rx+2.0, 7.8) {\scriptsize 8+8 CU-tasks};

\node[xcd] (f-oproj) at (\rx+1.0, 6.9) {O-Proj+Res};
\node[annot, anchor=west] at (\rx+2.0, 6.9) {\scriptsize 8 Chiplet-tasks};

\node[cu] (f-rms2) at (\rx+1.0, 6.0) {RMSNorm};
\node[annot, anchor=west] at (\rx+2.0, 6.0) {\scriptsize 1 CU-task};

\node[xcd] (f-gate) at (\rx+1.0, 5.1) {Gate+Up+SiLU};
\node[annot, anchor=west] at (\rx+2.0, 5.1) {\scriptsize 8 Chiplet-tasks};

\node[xcd] (f-down) at (\rx+1.0, 3.3) {Down+Res};
\node[annot, anchor=west] at (\rx+2.0, 3.3) {\scriptsize 8 Chiplet-tasks};

\draw[arr] (f-rms1) -- (f-qkv);
\draw[arr] (f-qkv) -- (f-attn);
\draw[arr] (f-attn) -- (f-oproj);
\draw[arr] (f-oproj) -- (f-rms2);
\draw[arr] (f-rms2) -- (f-gate);
\draw[arr] (f-gate) -- (f-down);

\draw[thick, dashed, gray!30] (4.5, 2.8) -- (4.5, 10.6);

\node[font=\scriptsize, text=black, align=center] at (1.0, 2.6) {No XCD awareness\\96--256 tasks per GEMM};
\node[font=\scriptsize, text=black, align=center] at (\rx+1.0, 2.6) {8 Chiplet-tasks per GEMM\\SiLU fused into Gate+Up};

\def\px{13.0}
\node[font=\normalsize\bfseries, anchor=south] at (\px+3.0, 10.8) {(b) Runtime Dispatch on 2 XCDs};

\node[xcdbox, minimum width=9.0cm, minimum height=3.6cm] (xcd0box) at (\px+3.0, 8.8) {};
\node[font=\small\bfseries, black, anchor=north east] at (xcd0box.north east) {XCD 0};

\node[font=\scriptsize\bfseries, black, anchor=west] at (\px-1.2, 10.4) {CU-task (round-robin)};

\node[sched, minimum width=1.2cm] (s0) at (\px+0.0, 9.8) {Sched};

\node[wkr, fill=green!25, thick] (w0) at (\px+1.8, 9.8) {W\textsubscript{0}};
\node[wkr] (w1) at (\px+2.7, 9.8) {W\textsubscript{1}};
\node[wkr] (w2) at (\px+3.6, 9.8) {W\textsubscript{2}};
\node[font=\small, gray] at (\px+4.5, 9.8) {\ldots};
\node[wkr] (w30) at (\px+5.5, 9.8) {W\textsubscript{30}};

\draw[->, >=Stealth, very thick, green!60!black] (s0.east) -- (w0.west)
    node[midway, above, font=\tiny, green!50!black] {1 task};

\node[draw, rounded corners=2pt, minimum width=5.5cm, minimum height=0.45cm, fill=gray!8, font=\tiny] (q0) at (\px+3.6, 8.7) {per-worker task queues (global memory)};
\draw[arr, black!30] (w0.south) -- ([xshift=-1.5cm]q0.north);
\draw[arr, black!30] (w2.south) -- ([xshift=0cm]q0.north);
\draw[arr, black!30] (w30.south) -- ([xshift=1.5cm]q0.north);

\node[draw, rounded corners=2pt, minimum width=7.5cm, minimum height=0.45cm, fill=green!15, font=\scriptsize\bfseries] (l20) at (\px+3.0, 7.6) {4\,MB L2 (shared by all workers on XCD 0)};

\node[xcdbox, minimum width=9.0cm, minimum height=3.6cm] (xcd1box) at (\px+3.0, 4.2) {};
\node[font=\small\bfseries, black, anchor=north east] at (xcd1box.north east) {XCD 1};

\node[font=\scriptsize\bfseries, black, anchor=west] at (\px-1.2, 5.80) {Chiplet-task (broadcast)};

\node[sched, minimum width=1.2cm] (s1) at (\px+0.0, 5.2) {Sched};

\node[wkr, fill=blue!30, thick] (w31) at (\px+1.8, 5.2) {W\textsubscript{31}};
\node[wkr, fill=blue!30, thick] (w32) at (\px+2.7, 5.2) {W\textsubscript{32}};
\node[wkr, fill=blue!30, thick] (w33) at (\px+3.6, 5.2) {W\textsubscript{33}};
\node[font=\small, gray] at (\px+4.5, 5.2) {\ldots};
\node[wkr, fill=blue!30, thick] (w61) at (\px+5.5, 5.2) {W\textsubscript{61}};

\draw[very thick, blue!60!black] (s1.east) -- (\px+1.3, 5.2);
\draw[very thick, blue!60!black] (\px+1.3, 5.2) -- (\px+1.3, 5.55);
\draw[very thick, blue!60!black] (\px+1.3, 5.55) -- (\px+5.5, 5.55);
\draw[->, >=Stealth, thick, blue!60!black] (\px+1.8, 5.55) -- (w31.north);
\draw[->, >=Stealth, thick, blue!60!black] (\px+2.7, 5.55) -- (w32.north);
\draw[->, >=Stealth, thick, blue!60!black] (\px+3.6, 5.55) -- (w33.north);
\draw[->, >=Stealth, thick, blue!60!black] (\px+5.5, 5.55) -- (w61.north);

\node[draw, rounded corners=2pt, minimum width=5.5cm, minimum height=0.45cm, fill=gray!8, font=\tiny] (q1) at (\px+3.6, 4.1) {per-worker task queues (global memory)};
\draw[arr, black!30] (w31.south) -- ([xshift=-1.5cm]q1.north);
\draw[arr, black!30] (w33.south) -- ([xshift=0cm]q1.north);
\draw[arr, black!30] (w61.south) -- ([xshift=1.5cm]q1.north);

\node[draw, rounded corners=2pt, minimum width=7.5cm, minimum height=0.45cm, fill=green!15, font=\scriptsize\bfseries] (l21) at (\px+3.0, 3.0) {4\,MB L2 (shared by all workers on XCD 1)};

\node[draw, rounded corners=3pt, minimum width=1.8cm, minimum height=1.2cm, fill=red!10, font=\scriptsize\bfseries, align=center, text width=1.5cm] (gc) at (\px+8.5, 5.3) {Global\\events\\(HBM)};

\draw[arr, red!50, thick] (l20.east) -| (gc.north)
    node[pos=0.25, above, font=\tiny, black] {\texttt{buffer\_wbl2}};

\draw[arr, red!50, thick] (gc.south) |- (l21.east)
    node[pos=0.75, below, font=\tiny, black] {sched polls};

\end{tikzpicture}%
}
\caption{\sysname system overview. (a)~Task graph for one transformer layer at bs=1: standard dispatch decomposes each GEMM into 96--256 independent CU-tasks (1,407 total); \sysname uses eight Chiplet-tasks per GEMM (543 total, 2.6$\times$ fewer), with SiLU fused into the gate+up Chiplet-task. (b)~Runtime dispatch: each XCD has a scheduler that assigns CU-tasks round-robin to individual workers, or broadcasts Chiplet-tasks to all workers for cooperative execution with strided tiles. Task queues and event counters reside in global memory (HBM) but are naturally cached in the XCD-local L2; intra-XCD communication is implicitly coherent through the shared L2 without global acquire/release semantics. Cross-XCD signaling requires an explicit \texttt{buffer\_wbl2} fence; schedulers poll the global event counter to detect upstream completion.}
\label{fig:system-overview}
\end{figure*}

\subsection{Persistent Kernel Runtime}
\label{sec:runtime}
\sysname extends the Mirage Persistent Kernel system~\cite{Cheng:2025:MPK} with chiplet-aware concepts and hierarchical synchronization.  To implement this, our runtime launches a single CUDA/HIP kernel that occupies every CU on the GPU: one workgroup per XCD serves as a scheduler that dispatches work with L2-local affinity, while the rest serve as workers executing tasks.  Each scheduler discovers its XCD identity by reading the hardware \texttt{HW\_ID} register and maintains per-worker task queues in global memory.
As a result, CUs occupied by scheduler blocks (8 of 256 CUs (3.1\%) in MI350) are unavailable for computation. However, this allows schedulers to resolve event dependencies and to enqueue the next task concurrently with worker execution.
Because schedulers perform only lightweight control operations (counter reads, queue writes, pointer arithmetic), we observe that workers rarely stall waiting for task assignment in our designs.  However, in designs with very short executing tasks, scheduling overhead might become more significant.

\subsection{Hierarchical Synchronization}
\label{sec:two-level-events}

Since \sysname replaces hardware-based work-dispatch mechanisms with scheduler threads running on CUs, synchronization costs directly impact dispatch throughput. On monolithic GPUs, device-scope atomics and fences are relatively cheap, all CUs share a single L2 cache, so coherence traffic stays on-chip. However, on chiplet architectures like MI350X, a device-scope atomic must be visible across all eight XCDs, each with its own 4 MB L2 partition. This requires coherence traffic over the inter-chiplet fabric, inflating atomic latency compared to an L2-local operation. Similarly, a device-scope fence (buffer\_wbl2) flushes dirty cachelines from the local L2 to ensure cross-XCD visibility. This operation scales with the number of dirty lines and penalizes all subsequent memory accesses until the flush completes. To avoid these costs, \sysname uses a hierarchical synchronization scheme which matches each communication pattern to the narrowest memory scope sufficient for correctness:
  
\begin{enumerate}
\item \textbf{Task queue (read-only).} The global queue of task descriptors is populated before the megakernel launches and remains immutable during execution. Because no writer contention exists, workers and schedulers read task metadata without any synchronization.
\item \textbf{Scheduler $\to$ worker dispatch (L2-local).} Each scheduler assigns tasks to workers on its own XCD by writing to per-worker queues using device-scope stores and atomics. Since both the scheduler and its workers reside on the same XCD, these accesses resolve in the local L2 cache without cross-XCD coherence traffic.
\item \textbf{Worker $\to$ worker coordination (L2-local).} Workers within a Chiplet-task accumulate sub-task completion counts in per-XCD counters using device-scope atomics, which also resolve in the local L2. No fence is required because all participating workers share the same L2 partition.
\item \textbf{XCD $\to$ global signaling (GPU-scope).} Only the last worker to complete on each XCD issues a single \texttt{threadfence} and updates the global event counter using a GPU-scope atomic (\texttt{flat\_atomic\_add} with \texttt{sc0 sc1}). This amortizes the expensive cross-XCD coherence cost across all workers in the XCD. Once the global counter reaches its threshold, the completing worker enqueues the event to the responsible scheduler's queue to trigger downstream tasks.
\end{enumerate}


\begin{figure}[t]
\centering
\resizebox{\columnwidth}{!}{%
\begin{tikzpicture}[
    worker/.style={draw, rounded corners=2pt, minimum width=0.8cm, minimum height=0.55cm, fill=blue!15, font=\tiny\bfseries, inner sep=1pt},
    localctr/.style={draw, rounded corners=3pt, minimum width=2.4cm, minimum height=0.6cm, fill=green!20, font=\scriptsize\bfseries, thick},
    globalctr/.style={draw, rounded corners=3pt, minimum width=3.0cm, minimum height=0.6cm, fill=red!15, font=\scriptsize\bfseries, thick},
    xcdbox/.style={draw, dashed, rounded corners=5pt, inner sep=8pt, thick, blue!40},
    l2label/.style={font=\tiny, text=green!50!black},
    hbmlabel/.style={font=\tiny, text=red!60!black},
    annot/.style={font=\tiny, text=black!70, align=center},
    bigarrow/.style={->, >=Stealth, thick},
    fastarrow/.style={->, >=Stealth, green!60!black, thick},
    slowarrow/.style={->, >=Stealth, red!60!black, very thick},
    >=Stealth
]

\node[xcdbox, minimum width=4.2cm, minimum height=4.0cm] (xcd0box) at (0, 0) {};
\node[font=\scriptsize\bfseries, blue!60] at (0, 1.75) {XCD 0 (L2 coherent)};

\node[worker] (w00) at (-1.3, 0.9) {W\textsubscript{0}};
\node[worker] (w01) at (-0.4, 0.9) {W\textsubscript{1}};
\node[worker] (w02) at (0.5, 0.9) {\ldots};
\node[worker] (w03) at (1.3, 0.9) {W\textsubscript{30}};

\node[localctr] (lc0) at (0, -0.4) {$L[0][e]$};

\draw[fastarrow] (w00.south) -- ++(0, -0.35) -| ([xshift=-8pt]lc0.north);
\draw[fastarrow] (w01.south) -- ++(0, -0.2) -| ([xshift=-3pt]lc0.north);
\draw[fastarrow] (w02.south) -- ++(0, -0.2) -| ([xshift=3pt]lc0.north);
\draw[fastarrow] (w03.south) -- ++(0, -0.35) -| ([xshift=8pt]lc0.north);

\node[annot, green!50!black, anchor=east] at (-1.5, 0.2) {no fence};

\node[font=\tiny, red!70!black, anchor=west] at (1.65, 0.9) {last?};

\node[xcdbox, minimum width=4.2cm, minimum height=4.0cm] (xcd1box) at (6.0, 0) {};
\node[font=\scriptsize\bfseries, blue!60] at (6.0, 1.75) {XCD 1 (L2 coherent)};

\node[worker] (w10) at (4.7, 0.9) {W\textsubscript{31}};
\node[worker] (w11) at (5.6, 0.9) {W\textsubscript{32}};
\node[worker] (w12) at (6.5, 0.9) {\ldots};
\node[worker] (w13) at (7.3, 0.9) {W\textsubscript{61}};

\node[localctr] (lc1) at (6.0, -0.4) {$L[1][e]$};

\draw[fastarrow] (w10.south) -- ++(0, -0.35) -| ([xshift=-8pt]lc1.north);
\draw[fastarrow] (w11.south) -- ++(0, -0.2) -| ([xshift=-3pt]lc1.north);
\draw[fastarrow] (w12.south) -- ++(0, -0.2) -| ([xshift=3pt]lc1.north);
\draw[fastarrow] (w13.south) -- ++(0, -0.35) -| ([xshift=8pt]lc1.north);

\node[annot, green!50!black, anchor=west] at (7.5, 0.2) {no fence};
\node[font=\tiny, red!70!black, anchor=west] at (7.65, 0.9) {last?};

\draw[thick, dashed, gray!50] (3.0, 2.2) -- (3.0, -3.1);
\node[font=\tiny, gray, rotate=90, anchor=south] at (3.35, 0.8) {XCD boundary};

\node[draw, rounded corners=2pt, minimum width=9.4cm, minimum height=0.45cm, fill=orange!10, thick, font=\scriptsize] (fence) at (3.0, -1.7) {\texttt{buffer\_wbl2} fence (L2 $\rightarrow$ HBM writeback)};

\draw[slowarrow] (lc0.south) -- ([xshift=-1.5cm]fence.north)
    node[midway, left, font=\tiny, red!60!black] {$L{=}T$?};
\draw[slowarrow] (lc1.south) -- ([xshift=1.5cm]fence.north)
    node[midway, right, font=\tiny, red!60!black] {$L{=}T$?};

\node[globalctr] (gc) at (3.0, -2.5) {Global $G[e]$};
\node[hbmlabel] at (3.0, -2.95) {\texttt{flat\_atomic\_add sc0 sc1}};

\draw[slowarrow] ([xshift=-1cm]fence.south) -- ([xshift=-0.5cm]gc.north);
\draw[slowarrow] ([xshift=1cm]fence.south) -- ([xshift=0.5cm]gc.north);

\node[draw, rounded corners=3pt, minimum width=2.4cm, minimum height=0.5cm, fill=purple!12, font=\scriptsize\bfseries] (sched) at (3.0, -3.7) {Schedulers poll $G[e]$};
\draw[bigarrow, purple!60] (gc.south) -- (sched.north)
    node[midway, right, font=\tiny, purple!60!black] {$G \geq$ needed?};

\node[annot] (dispatch) at (3.0, -4.4) {dispatch downstream tasks};
\draw[bigarrow, purple!60] (sched.south) -- (dispatch.north);


\end{tikzpicture}%
}
\caption{Hierarchical synchronization protocol. Workers increment XCD-local L2 counters without fences (green). Only the last worker on each XCD issues a single \texttt{buffer\_wbl2} fence and updates the global HBM counter (red). Schedulers poll the global counter to dispatch downstream tasks. This reduces cross-XCD fences compared to per-worker global signaling.}
\label{fig:hierarchical-sync}
\end{figure}

Using these fine-grained scopes is more complicated, but generally more efficient, than generic fence operations. Device-scope fences such as CUDA/HIP's \texttt{\_\_threadfence} or \texttt{\_\_builtin\_amdgcn\_fence} with \texttt{\_\_ATOMIC\_RELEASE} and ``agent'' scope generally lower to \texttt{buffer\_wbl2}, which flushes dirty cachelines from the L2 cache. For wavefront-tasks and CU-tasks that execute on a single CU, workers signal completion directly via a GPU-scope atomic operation on the global event counter; no two-level counting is needed, since there is only one worker per task.

For Chiplet-tasks, two-level counting issues at most one fence per XCD per event, and zero fences on XCDs with no tasks for that event. \sysname linear events have exactly eight tasks (one per XCD), so each triggers exactly eight fences total. By using L2-local counters for intra-chiplet coordination, two-level counting avoids per-worker global fences entirely, as shown in Figure~\ref{fig:hierarchical-sync}.
  
\subsection{Task Registration and Code Generation}
\label{sec:codegen}

\sysname extends the Mirage\cite{Cheng:2025:MPK} pipeline with \sysname-specific library code and new node descriptors for each Chiplet-task. The Mirage compiler generates GPU kernel code at model load time by compiling an abstract task graph into a runtime description that includes library code for each node. To generate a task graph based on Chiplet-tasks, we currently provide different input code to the Mirage compiler and do not rely on compiler-driven super-optimization. We leave compiler-driven super-optimization as future work, noting that it requires only an appropriate cost model for Chiplet-tasks.

When operations are decomposed into Chiplet-tasks, the resulting task graph typically contains eight Chiplet-tasks for the operation, one for each physical XCD.  We rely on the Mirage compiler to generate pointers for each tile at code generation time. (i.e., for a weight partition across eight XCDs, task $k$ receives a pointer which is $\text{base\_ptr} + k \times (N/8) \times K \times \text{sizeof(bf16)}$).  As a result, GPU kernel code does not typically need to consider which Chiplet-task it is executing; it simply reads a pointer to the current tile of data.  However, within each Chiplet-task, each CU receives the same pointer, which avoids the scheduler needing to perform pointer arithmetic at task dispatch time.  Instead, we modify the call emitted from the Mirage compiler to include the tile index within the XCD, generating a call to \texttt{\_execute\_xcd\_task(tile\_idx)} instead of \texttt{\_execute\_task()}.  The implementation of the library is responsible for implementing different behaviors for each CU based on the tile index.

\section{End-to-end Evaluation}
\label{sec:eval}
\subsection{Experimental Setup}

We perform evaluations on a single GPU AMD Instinct MI350X system.  This device has 256 CUs across eight XCDs (32 CUs/XCD), 4\,MB L2 per XCD, 256\,MB shared LLC, and 288 GB HBM3.
We compare against 2 baseline implementations:

(1)~Mirage MPK~\cite{Cheng:2025:MPK} ported to AMD MI350X: a chiplet-unaware persistent kernel using standard 2D GEMM tiling with no cooperative scheduling, serving as the \emph{internal baseline} which isolates the effect of \sysname's chiplet-aware optimizations;
(2)~vLLM~\cite{Kwon:2023:EMM} v0.17.2 on ROCm (\texttt{--enforce-eager}): standard per-operator kernel launches without persistent kernel, serving as the \emph{external baseline} representing the state-of-the-art serving framework.

Performance metrics were collected using a combination of HIP-level timing APIs, on-GPU cycle counters, performance counters and low-level tracing.
All end-to-end latency results use 64 input tokens and 1024 output tokens, with decode-only TPOT (Time Per Output Token) reported.
Decode latency is measured via GPU-side per-iteration timestamps (\texttt{s\_memrealtime}~\cite{AMD:2025:CDNA4ISA}\footnote{A timer inside the kernel that gives more accurate measurements.}), excluding prefill.

\subsection{End-to-End Decode Latency}

\begin{figure}[t]
\centering
\begin{tikzpicture}
\begin{axis}[
    xlabel={Batch Size},
    ylabel={Decode Latency (ms/token)},
    xtick={1,2,4,8,16,32,64},
    xmode=log,
    log basis x=2,
    ymin=0, ymax=32,
    legend style={at={(0.02,0.98)}, anchor=north west, font=\footnotesize, row sep=-1pt},
    width=\columnwidth,
    height=6.5cm,
    mark size=2.5pt,
    thick,
]
\addplot[blue, mark=square*, dashed] coordinates {(1,10.511) (2,10.464) (4,10.619) (8,12.626) (16,12.559) (32,11.579) (64,11.806)};
\addplot[gray, mark=diamond*, dotted] coordinates {(1,7.832) (2,8.499) (4,9.013) (8,9.664) (16,10.648) (32,15.621) (64,24.098)};
\addplot[red, mark=triangle*, thick] coordinates {(1,6.824) (2,7.375) (4,7.941) (8,8.472) (16,9.214) (32,12.348) (64,18.610)};
\addplot[orange, mark=pentagon*, densely dashdotted] coordinates {(1,6.734) (2,7.314) (4,7.898) (8,8.505) (16,9.117) (32,13.370) (64,23.404)};
\legend{vLLM, Mirage, \sysname (M-tile), \sysname (M-split)}
\end{axis}
\end{tikzpicture}

\caption{Decode-only TPOT (ms/token) for Qwen3-8B (dense, bf16) on MI350.
\sysname (M-tile) uses an M-major windowed traversal, in which consecutive workers share weight tiles in L2.
\sysname (M-split) assigns each XCD a disjoint M-tile with no cross-worker weight sharing. Both use 16x64x256 CK GEMM tiles and cache-streaming weight loads.}
\label{fig:latency}
\end{figure}

We evaluate \sysname on a dense Qwen3-8B~\cite{Qwen:2025:QTR} model (4096 hidden dim, 36 layers, 12288 FFN dim, GQA with 32 query / 8 KV heads) with bf16.

Figure~\ref{fig:latency} compares four configurations:
(1)~vLLM~\cite{Kwon:2023:EMM} using hipblaslt/Tensile GEMMs with kernel-per-operator dispatch;
(2)~Mirage MPK~\cite{Cheng:2025:MPK} ported to MI350 with hierarchical scheduling but no chiplet awareness, using individual CU-tasks with no cooperative weight sharing;
(3)~\sysname (M-tile), which uses M-major windowed traversal so that consecutive workers within each XCD share weight tiles in L2; and
(4)~\sysname (M-split), which assigns each XCD a disjoint slice of the M (batch) dimension so that no two XCDs share weight tiles, isolating the scheduling benefit from the L2 reuse benefit.

The chiplet-unaware Mirage baseline is already 1.34$\times$ faster than vLLM at bs=1 (7.83 vs.\ 10.51\,ms) due to persistent kernel launch overhead elimination.
\sysname's hierarchical Chiplet-task scheduling reduces this further to 6.82\,ms (M-tile) and 6.73\,ms (M-split), which is 1.54$\times$ and 1.56$\times$ faster than vLLM, respectively.
At small batch sizes (bs=1--16), where $m\_tiles=1$ and no cooperative weight reuse occurs, both \sysname variants perform similarly because the improvement comes entirely from reduced scheduling overhead: eight Chiplet-tasks per GEMM versus 96--256 CU-tasks in the Mirage baseline.

The two \sysname variants diverge at bs$\geq$32, where $m\_tiles \geq 2$ and cooperative weight sharing activates in M-tile mode.
At bs=32, M-tile achieves 51.0\% L2 hit rate versus 39.5\% for M-split and 38.9\% for Mirage (Table~\ref{tab:l2}), translating to 12.35\,ms versus 13.37\,ms and 15.62\,ms, respectively.
At bs=64, the gap widens further: M-tile reaches 61.4\% L2 hit rate and 18.61\,ms, while M-split falls to 47.4\% and 23.40\,ms, close to Mirage's 24.10\,ms, demonstrating that L2 cooperative tiling is the dominant factor at large batch sizes.

At bs$\geq$32, vLLM's latency flattens at ${\sim}$11--12\,ms as hipBLASLt's GEMM kernels (with wave-level K-splitting, \S\ref{sec:nsplit-vs-ksplit}) efficiently exploit the compute-bound regime.
\sysname (M-tile) remains 1.06$\times$ faster at bs=32 but is slower at bs=64, as the persistent kernel's GEMM implementation does not yet include K-splitting optimizations for the compute-bound regime and attention optimizations.

\paragraph{Decomposing the speedup: scheduling vs.\ L2 locality.}
The M-split variant provides a clean ablation: it uses Chiplet-task scheduling
(eight tasks per GEMM) but assigns each XCD a disjoint M-tile, isolating the
scheduling benefit from L2 cooperative weight reuse.

At batch sizes 1--16, $m\_tiles=1$ and no cooperative weight reuse occurs.
Both M-tile and M-split perform similarly (1.13--1.16$\times$ faster than Mirage),
and Table~\ref{tab:l2} confirms that L2 hit rates and HBM traffic are nearly
identical across all three configurations (${\sim}$16--24\% L2 hit rate,
${\sim}$2400--2800\,GB HBM reads). The speedup at these batch sizes comes from reduced scheduler-to-worker communication overhead: each Chiplet-task requires a single dispatch from the scheduler to the workers on that XCD, whereas the standard decomposition issues many CU-level dispatches per GEMM.

At bs=32, $m\_tiles=2$ and the two variants diverge. M-tile, where consecutive
workers share weight columns in L2, reaches 51.0\% L2 hit rate and 1.27$\times$
speedup over Mirage (12.35 vs.\ 15.62\,ms). M-split, without weight sharing,
achieves only 39.5\% L2 hit rate and 1.17$\times$ speedup (13.37\,ms), consistent
with the scheduling-only improvement at smaller batch sizes. The gap between
them is directly attributable to L2 weight reuse, confirmed by HBM traffic:
M-tile reads 3,190\,GB versus Mirage's 3,906\,GB (18\% reduction), while
M-split shows no reduction (4,295\,GB, Table~\ref{tab:l2}).

At bs=64 ($m\_tiles=4$), four workers share each weight column and the L2
benefit dominates. M-tile reaches 61.4\% L2 hit rate and 1.30$\times$ speedup
(18.61 vs.\ 24.10\,ms), reducing HBM reads from 6,203\,GB to 3,925\,GB (37\%).
M-split falls to 1.03$\times$ (23.40\,ms), nearly identical to Mirage, with no HBM
reduction (7,426\,GB).                  

\subsection{Roofline Analysis}
\label{sec:roofline}

\begin{figure}[t]
\centering
\begin{tikzpicture}
\begin{axis}[
    xlabel={Operational Intensity (FLOP/byte)},
    ylabel={Attainable Performance (TFLOPS)},
    xmode=log,
    ymode=log,
    xmin=0.3, xmax=600,
    ymin=1, ymax=1800,
    width=\columnwidth,
    height=6cm,
    thick,
    legend style={at={(0.02,0.98)}, anchor=north west, font=\footnotesize, row sep=-2pt, fill=white, fill opacity=0.9, draw opacity=1, text opacity=1},
    clip=false,
]
\addplot[gray, thick, domain=0.3:245, forget plot] {5.3 * x};
\addplot[gray, thick, domain=245:600, forget plot] {1300};
\node[font=\tiny, gray, anchor=north west] at (axis cs:245, 1300) {ridge (245)};
\addplot[only marks, mark=square*, blue, mark size=3pt] coordinates {
    (1, 5.3)
    (8, 42.4)
    (16, 84.8)
    (32, 169.6)
};
\addplot[only marks, mark=triangle*, red, mark size=3pt] coordinates {
    (1.20, 6.4)
    (10.10, 53.5)
    (20.97, 111.1)
    (65.3, 346.1)
};
\draw[->, thick, red!60, densely dashed] (axis cs:1, 5.3) -- (axis cs:1.20, 6.4);
\draw[->, thick, red!60, densely dashed] (axis cs:8, 42.4) -- (axis cs:10.10, 53.5);
\draw[->, thick, red!60, densely dashed] (axis cs:16, 84.8) -- (axis cs:20.97, 111.1);
\draw[->, thick, red!60, densely dashed] (axis cs:32, 169.6) -- (axis cs:65.3, 346.1);

\node[font=\tiny, anchor=east] at (axis cs:0.9, 5.3) {bs=1};
\node[font=\tiny, anchor=east] at (axis cs:7, 42.4) {bs=8};
\node[font=\tiny, anchor=east] at (axis cs:14, 84.8) {bs=16};
\node[font=\tiny, anchor=east] at (axis cs:28, 169.6) {bs=32};

%
%
%
%
\addplot[mark=square, blue, mark size=2.5pt, only marks] coordinates {
    (6.8,1.5) (13.1,2.9) (25.5,4.5) (48.3,8.0) (91.5,17.7) (100.2,23.7)
};
\addplot[mark=triangle, red, mark size=2.5pt, only marks] coordinates {
    (6.5,1.7) (12.9,2.7) (24.5,5.2) (46.2,9.6) (84.0,18.3) (127.6,36.6)
};
\node[font=\tiny, anchor=south] at (axis cs:6.5, 1.9) {1};
\node[font=\tiny, anchor=west] at (axis cs:10.5, 5.7) {2};
\node[font=\tiny, anchor=south] at (axis cs:20.5, 6.8) {4};
\node[font=\tiny, anchor=west] at (axis cs:24, 14.6) {8};
\node[font=\tiny, anchor=south] at (axis cs:50.0, 18.5) {16};
\node[font=\tiny, anchor=south] at (axis cs:80.6, 38.5) {32};
\node[font=\tiny, gray, rotate=38, anchor=south] at (axis cs:2, 14) {HBM: 5.3\,TB/s};

\legend{Standard (theoretical), \sysname (theoretical), Mirage (measured), \sysname (measured)}
\end{axis}
\end{tikzpicture}
\caption{Roofline analysis of GEMM operations on MI350 (bf16 MFMA). Standard scheduling operates at nominal arithmetic intensity $\text{AI} = B$ (batch size in FLOP/byte). \sysname's L2 reuse increases effective AI by reducing HBM traffic: $\text{AI}_\text{eff} = B / (1 - \text{L2\ hit\ rate})$. At bs=32 with 51\% L2 hit rate, \sysname shifts the effective AI from 32 to 65, a 2.0$\times$ rightward shift toward the ridge point (245).}
\label{fig:roofline}
\end{figure}

Figure~\ref{fig:roofline} presents a roofline analysis of GEMM operations on MI350. Theoretical points (filled markers) plot each batch size at its nominal arithmetic intensity $\text{AI} = B$; \sysname's L2 reuse increases the effective intensity to $\text{AI}_\text{eff} = B / (1 - \text{L2 hit rate})$, shifting operations rightward toward the compute-bound regime. At bs=32, cooperative tiling increases effective intensity from 32 to 65 FLOP/byte. Measured points (open markers), obtained from hardware performance counters, corroborate this: \sysname achieves higher effective intensity than Mirage at each batch size, with the difference growing as L2 reuse increases. Table~\ref{tab:l2} reports the corresponding L2 hit rates and HBM traffic.

\subsection{L2 Hit Rate Model}

\begin{table*}[t]
\centering
\caption{L2 cache and HBM traffic for Qwen3-8B on MI350 (1024 decode tokens). HBM Rd/Wr are normalized to Mirage at each batch size (1.00$\times$). Measured via \texttt{rocprofiler-sdk} device counting mode.}
\label{tab:l2}
\small
\resizebox{\textwidth}{!}{%
\begin{tabular}{l*{4}{r}*{3}{r}*{3}{r}}
\toprule
& \multicolumn{4}{c}{\textbf{L2 Hit\%}} & \multicolumn{3}{c}{\textbf{HBM Rd ($\times$ Mirage)}} & \multicolumn{3}{c}{\textbf{HBM Wr ($\times$ Mirage)}} \\
\cmidrule(lr){2-5} \cmidrule(lr){6-8} \cmidrule(lr){9-11}
\textbf{BS} & \textbf{Mirage} & \textbf{NoSplitK} & \shortstack{\textbf{\sysname}\\\textbf{(M-tile)}} & \shortstack{\textbf{\sysname}\\\textbf{(M-split)}} & \textbf{NoSplitK} & \shortstack{\textbf{\sysname}\\\textbf{(M-tile)}} & \shortstack{\textbf{\sysname}\\\textbf{(M-split)}} & \textbf{NoSplitK} & \shortstack{\textbf{\sysname}\\\textbf{(M-tile)}} & \shortstack{\textbf{\sysname}\\\textbf{(M-split)}} \\
\midrule
1  & 16.4\% & 23.4\% & 16.9\% & 17.0\% & 0.98 & 0.98 & 0.99 & 0.29 & 0.32 & 0.87 \\
2  & 17.3\% & 24.2\% & 17.5\% & 17.9\% & 1.01 & 1.01 & 1.02 & 0.71 & 0.40 & 0.76 \\
4  & 18.2\% & 25.1\% & 18.9\% & 19.2\% & 0.98 & 0.97 & 0.98 & 0.37 & 0.29 & 0.76 \\
8  & 20.0\% & 26.3\% & 20.8\% & 20.8\% & 1.01 & 1.00 & 1.01 & 0.60 & 0.33 & 0.38 \\
16 & 22.2\% & 28.7\% & 23.7\% & 23.5\% & 0.99 & 0.97 & 0.98 & 0.47 & 0.45 & 0.41 \\
32 & 38.9\% & 42.6\% & 51.0\% & 39.5\% & 1.04 & 0.82 & 1.10 & 0.45 & 0.38 & 0.40 \\
64 & 39.0\% & 41.9\% & 61.4\% & 47.4\% & 1.08 & 0.63 & 1.20 & 0.48 & 0.44 & 0.44 \\


\bottomrule
\end{tabular}%
}
\end{table*}

The L2 hit rate from cooperative tiling can be modeled analytically. Under M-major traversal, $R = \min(W, m\_tiles)$ workers process the same weight tile before advancing to the next column, yielding:
\begin{equation}
\text{L2 Hit}_{\text{weight}} = (R - 1)/R = 1 - 1/\min(W, m\_tiles)
\label{eq:l2-hit}
\end{equation}
At bs=1 ($R{=}1$), the model predicts zero weight reuse, yet we measure 16.9\%.
This baseline L2 hit rate comes from \emph{fused SiLU}: the gate and up projections are fused into a single GEMM over a concatenated $[W_{\text{gate}}; W_{\text{up}}]$ weight matrix, so the two halves share activation reads and generate L2 hits during the fused operation.
Without fused SiLU, the measured L2 hit rate at bs=1 drops to ${\sim}$9\%.
At larger batch sizes, L2 hits serve data at aggregate L2 bandwidth ($\sim$100\,TB/s across eight XCDs) rather than HBM (5.3\,TB/s), amplifying effective memory bandwidth.

\paragraph{Per-GEMM weight analysis.}
\begin{table}[h]
\centering
\caption{Per-GEMM weight sizes for Qwen3-8B (bf16). L2 window = active working set per worker (one K-chunk tile).}
\label{tab:weight-sizes}
\small
\resizebox{\columnwidth}{!}{%
\begin{tabular}{lrrrr}
\toprule
\textbf{GEMM} & \textbf{Weight} & \textbf{Per-XCD} & \textbf{L2 window} & \textbf{Fits MALL?} \\
\midrule
qkv\_proj ($4096{\times}6144$)    & 48\,MB  & 6\,MB   & 384\,KB  & Yes \\
o\_proj ($4096{\times}4096$)      & 32\,MB  & 4\,MB   & 256\,KB  & Yes \\
gate\_up ($4096{\times}24576$)    & 192\,MB & 24\,MB  & 1.5\,MB  & Yes \\
down\_proj ($12288{\times}4096$)  & 96\,MB  & 12\,MB  & 256\,KB  & Yes \\
\midrule
\textbf{All 4 / layer} & \textbf{368\,MB} & \textbf{46\,MB} & --- & \textbf{No} (1.4$\times$) \\
\bottomrule
\end{tabular}%
}
\end{table}

Table~\ref{tab:weight-sizes} shows per-GEMM weight sizes. While no GEMM's per-XCD partition fits in the 4\,MB L2, cooperative tiling ensures that the \emph{active} working set, one K-chunk tile per worker, $31 \times 32\text{\,KB} \approx 1$\,MB remains resident. This is why M-major traversal achieves high hit rates despite per-XCD partitions exceeding L2 capacity by up to 6$\times$.
\section{Related Work}                                                            
\label{sec:related}
\paragraph{Mega-kernels for LLM inference.}
The HazyResearch mega-kernel~\cite{Spector:2025:LMN} fuses an entire Llama decoder into one persistent kernel with an on-GPU interpreter, achieving 78\% of H100 memory bandwidth at bs=1. FlashFormer~\cite{Nrusimha:2025:FWM} takes a similar whole-model fusion approach with pipelined shared buffers. Mirage MPK~\cite{Cheng:2025:MPK} introduces compiler-generated task graphs with event-driven synchronization, achieving 1.0--1.7$\times$ speedup over SGLang/vLLM on A100/H100. All target monolithic NVIDIA GPUs with a unified L2; \sysname addresses the orthogonal challenge of chiplet architectures with partitioned L2.

\paragraph{Cache-aware and chiplet-aware scheduling.}
Thread-block swizzling~\cite{Tillet:2019:TAI, Kerr:2017:CFL} and CUTLASS persistent GEMMs~\cite{NVIDIA:2024:CCT} improve L2 reuse but assume a unified cache. HipKittens~\cite{Hu:2025:HFA} ports ThunderKittens~\cite{Spector:2025:TSF} to AMD CDNA3/CDNA4 with XCD grouping, improving standalone GEMMs by up to 19\%. Both optimize individual kernel launches; \sysname extends chiplet awareness to a persistent mega-kernel that coordinates GEMM, attention, normalization, and activation tasks using two-level event synchronization.

\paragraph{Thread block clusters and topology-aware execution.}
NVIDIA's thread block clusters~\cite{NVIDIA:2024:PGT} (Hopper/Blackwell) co-schedule blocks on the same graphics processing clusters (GPC) with DSMEM sharing, cluster-scope barriers, and TMA multicast.  However, GPCs are a smaller granularity than a chiplet and lack a direct correspondence with the L2 level of memory hierarchy.  In contrast, \sysname's Chiplet-task model allows threads to coordinate across the entire L2 scope and can be efficiently implemented entirely in software.

\paragraph{Mirage superoptimizer.}
Mirage~\cite{Wu:2025:MAM} discovers optimal operator fusions via \emph{muGraphs}. \sysname builds on Mirage's MPK runtime~\cite{Cheng:2025:MPK} and adds chiplet awareness: where MPK dispatches tasks without regard for XCD boundaries, \sysname partitions weights across XCDs and coordinates workers via cooperative scheduling.
The two are complementary: Mirage decides \emph{what to fuse}; \sysname decides \emph{where to schedule}.

\paragraph{LLM serving systems.}
vLLM~\cite{Kwon:2023:EMM} (PagedAttention) and SGLang~\cite{Zheng:2023:SEE} (RadixAttention) are the dominant open-source serving frameworks. Both rely on CUDA/HIP kernel calls to low-level libraries, such as rocBLAS/hipBLASLt, which can implement some types of L2 optimization for core GEMM functions, but can make it difficult to express application-specific locality. \sysname provides a framework for arbitrary applications to leverage L2 locality, combined with a megakernel approach that avoids the runtime overhead of separate kernel invocations.




\section{Ablation Study}
\label{sec:appendix-ablation}

\paragraph{Generality across chiplet architectures.} 
The Chiplet-task abstraction is parameterized by chiplet count ($X$), workers per chiplet ($W$), and L2 capacity per chiplet ($C$). \sysname queries these at runtime, allowing the same task graph and scheduling logic to adapt to MI300X (8 XCDs, 38 CUs, 4\,MB L2) and MI350 (8 XCDs, 32 CUs, 4\,MB L2) without code changes. The abstraction itself is not AMD-specific: any chiplet GPU with partitioned L2 caches, including NVIDIA Blackwell~\cite{NVIDIA:2024:NBA}, could implement Chiplet-tasks with a platform-specific runtime backend. Existing serving frameworks (vLLM~\cite{Kwon:2023:EMM}, SGLang~\cite{Zheng:2023:SEE}) could adopt the abstraction incrementally; the key requirement is a software scheduler with task-to-chiplet mapping.
  
\paragraph{Interaction with tensor parallelism.}
In multi-GPU serving, tensor parallelism (TP) shards weight matrices across GPUs before \sysname partitions within a GPU.
\sysname operates on the local shard: if TP=4 reduces a GEMM's $N$ dimension by $4\times$, the per-XCD partition shrinks proportionally, potentially falling below the L2 capacity threshold and making cooperative dispatch unnecessary.
The two levels compose naturally because TP is an inter-GPU partitioning and \sysname scheduling is an intra-GPU partitioning, with no interaction in the synchronization path.
We leave multi-GPU evaluation to future work.

\paragraph{Register pressure and occupancy.}
The persistent megakernel compiles all task types into a single GPU function, and the combined register footprint limits occupancy to a single wave per SIMD, eliminating latency hiding from wave switching. Every L2 miss directly stalls the MFMA pipeline, making the L2 hit rate even more critical. A per-task compilation strategy could reduce register pressure but would sacrifice the single-launch property; we consider this an inherent tradeoff of the megakernel approach. Hardware support for dynamic register allocation, as implemented in AMD's RDNA architecture for graphics workloads, could mitigate this tradeoff by allowing each task type to consume only the registers it needs rather than the worst-case union.

\paragraph{Limitations and future work.}
Our evaluation covers one model (Qwen3-8B dense) on a single GPU architecture (MI350). We do not evaluate multi-GPU configurations, prefill performance, or models requiring tensor parallelism to fit in memory. An interesting direction is to integrate with MLIR-based kernel compilers to provide a unified Python-to-GPU stack in which chiplet-aware scheduling and tile-level codegen share a common IR.

\section{Conclusion}
Chiplet GPUs partition compute and caches across dies, creating a NUMA-like memory hierarchy that kernel-at-a-time execution cannot exploit. \sysname addresses this with a persistent megakernel and software scheduler that operates as a tiny OS and replaces the hardware block dispatcher, enabling two optimizations that are difficult to express in current programming models: cooperative weight tiling across workers within each XCD, and a hierarchical two-level event synchronization that confines most coherence traffic to the local L2.
On AMD Instinct MI350X with Qwen3-8B, \sysname achieves 1.3--1.5$\times$ lower per-token decode latency than vLLM across bs=1--16. At small batch sizes, the gain comes from reduced dispatch overhead via Chiplet-task scheduling. At bs=32, cooperative M-major weight tiling raises L2 hit rates from 38.9\% to 51.0\%, reducing HBM read traffic by 18\%. An M-split ablation, which uses Chiplet-task scheduling but no cooperative weight sharing, confirms that L2 reuse is the dominant source of speedup at larger batch sizes, while scheduling overhead reduction dominates at small batch sizes. These results demonstrate that software-managed scheduling can recover the L2 locality that chiplet architectures fragment, without requiring hardware changes.

\bibliographystyle{ACM-Reference-Format}
\bibliography{main}

\end{document}

%% file: figures/mem_hierarchy.tex
\begin{figure*}[t]
\centering
\includegraphics[width=\textwidth]{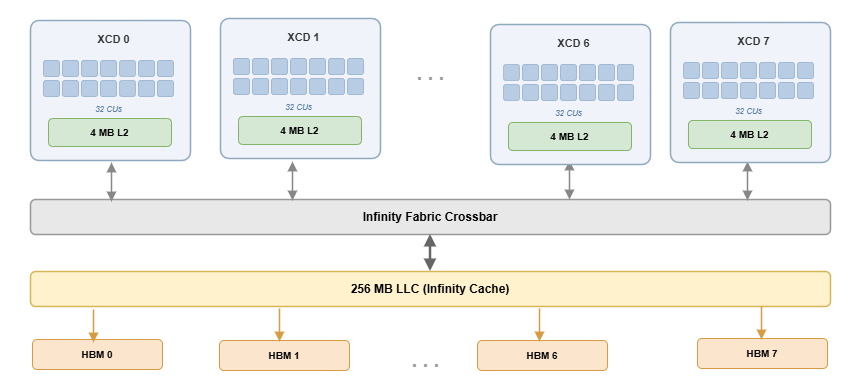}
\caption{AMD Instinct MI350 memory hierarchy. Each of 8 XCDs has 32 CUs and a private 4\,MB L2 cache (TCC). All XCDs share a 256\,MB MALL (Infinity Cache) before HBM3.}
\label{fig:mi300x-arch}
\end{figure*}

%% file: figures/xcd_tiling.tex
\definecolor{wc0}{RGB}{217,128,128}
\definecolor{wc1}{RGB}{217,166,115}
\definecolor{wc2}{RGB}{209,199,102}
\definecolor{wc3}{RGB}{115,191,128}
\definecolor{wc4}{RGB}{115,191,204}
\definecolor{wc5}{RGB}{140,140,217}

\begin{figure*}[t]
\centering
\resizebox{\textwidth}{!}{%
\begin{tikzpicture}[
    cell/.style={draw=black!40, rounded corners=2pt,
        minimum width=1.0cm, minimum height=0.85cm,
        inner sep=0pt},
    badge/.style={circle, fill=white, fill opacity=0.85,
        text opacity=1, inner sep=0pt, minimum size=0.5cm,
        font=\scriptsize\bfseries, text=black!75},
    >=Stealth
]

\node[font=\normalsize\bfseries, anchor=south] at (3.5, 6.6)
    {(a) N-major traversal};
\node[font=\small, anchor=south, text=black!50] at (3.5, 5.9)
    {Weight columns $[\text{K}, \text{N}/8]$};
\foreach \n in {0,...,5} {
    \node[font=\scriptsize\bfseries, text=wc\n] at (\n*1.1+0.55, 5.5) {n\n};
}
\node[font=\small, text=black!50, anchor=south, rotate=90] at (-1.1, 3.2) {Activations};
\foreach \m in {0,...,3} {
    \node[font=\scriptsize, text=black!50] at (-0.45, 4.65-\m*1.0) {m\m};
}
\foreach \m in {0,...,3} {
    \foreach \n in {0,...,5} {
        \pgfmathtruncatemacro{\idx}{\m*6+\n}
        \pgfmathtruncatemacro{\ts}{int(\idx/4)}
        \pgfmathtruncatemacro{\bright}{100 - 16*\ts}
        \colorlet{cellcol}{wc\n!\bright!black}
        \node[cell, fill=cellcol] at (\n*1.1+0.55, 4.65-\m*1.0) {};
    }
}
\foreach \i in {0,...,22} {
    \pgfmathtruncatemacro{\inext}{\i+1}
    \pgfmathtruncatemacro{\curm}{int(\i/6)}
    \pgfmathtruncatemacro{\curn}{mod(\i,6)}
    \pgfmathtruncatemacro{\nxtm}{int(\inext/6)}
    \pgfmathtruncatemacro{\nxtn}{mod(\inext,6)}
    \draw[black!30, line width=0.7pt, shorten >=0.25cm, shorten <=0.25cm]
        (\curn*1.1+0.55, 4.65-\curm*1.0) -- (\nxtn*1.1+0.55, 4.65-\nxtm*1.0);
}
\foreach \m in {0,...,3} {
    \foreach \n in {0,...,5} {
        \pgfmathtruncatemacro{\idx}{\m*6+\n}
        \node[badge] at (\n*1.1+0.55, 4.65-\m*1.0) {\idx};
    }
}

\def\rx{8.5}
\node[font=\normalsize\bfseries, anchor=south] at (\rx+3.5, 6.6)
    {(b) M-major traversal};
\node[font=\small, anchor=south, text=black!50] at (\rx+3.5, 5.9)
    {Weight columns $[\text{K}, \text{N}/8]$};
\foreach \n in {0,...,5} {
    \node[font=\scriptsize\bfseries, text=wc\n] at (\rx+\n*1.1+0.55, 5.5) {n\n};
}
\foreach \m in {0,...,3} {
    \foreach \n in {0,...,5} {
        \pgfmathtruncatemacro{\idx}{\n*4+\m}
        \pgfmathtruncatemacro{\ts}{int(\idx/4)}
        \pgfmathtruncatemacro{\bright}{100 - 16*\ts}
        \colorlet{cellcol}{wc\n!\bright!black}
        \node[cell, fill=cellcol] at (\rx+\n*1.1+0.55, 4.65-\m*1.0) {};
    }
}
\foreach \i in {0,...,22} {
    \pgfmathtruncatemacro{\inext}{\i+1}
    \pgfmathtruncatemacro{\curm}{mod(\i,4)}
    \pgfmathtruncatemacro{\curn}{int(\i/4)}
    \pgfmathtruncatemacro{\nxtm}{mod(\inext,4)}
    \pgfmathtruncatemacro{\nxtn}{int(\inext/4)}
    \draw[black!30, line width=0.7pt, shorten >=0.25cm, shorten <=0.25cm]
        (\rx+\curn*1.1+0.55, 4.65-\curm*1.0) -- (\rx+\nxtn*1.1+0.55, 4.65-\nxtm*1.0);
}
\foreach \m in {0,...,3} {
    \foreach \n in {0,...,5} {
        \pgfmathtruncatemacro{\idx}{\n*4+\m}
        \node[badge] at (\rx+\n*1.1+0.55, 4.65-\m*1.0) {\idx};
    }
}

\draw[thick, dashed, gray!30] (7.55, 1.2) -- (7.55, 7.0);

\end{tikzpicture}%
}
\caption{Tile traversal order within one \xcdtask (4$\times$6 output tile grid, 4 workers illustrated).
Cell hue indicates which weight column is accessed; darker shading indicates later timesteps.
(a)~N-major: consecutive tiles advance along weight columns, so concurrent workers load distinct weight data, increasing L2 cache pressure.
(b)~M-major: consecutive tiles advance down activation rows and share the same weight column, increasing temporal L2 cache locality.}
\label{fig:tiling}
\end{figure*}